\documentclass[a4paper,11pt,fleqn]{article}

\usepackage[ansinew]{inputenc}
\usepackage[mathscr]{eucal}
\usepackage{amsmath,amssymb,amsthm} 

\allowdisplaybreaks
\newcommand{\p}{\partial}

\newcommand{\todo}[1][\null]{\ensuremath{\clubsuit}}

\newcommand{\noprint}[1]{}
\newcommand{\checked}[1][\null]{\ensuremath{\boldsymbol{\surd}}}

\newtheorem{theorem}{Theorem}

\newtheorem{corollary}{Corollary}

{\theoremstyle{definition}

\newtheorem*{remark*}{Remark}
}

\begin{document}

\par\noindent {\LARGE\bf
Differential invariants for a class of diffusion equations
\par}

{\vspace{4mm}\par\noindent {\bf Elsa Dos Santos Cardoso-Bihlo$^\dag$, Alexander Bihlo$^\dag$ and Roman O.\ Popovych$^\ddag$
} \par\vspace{2mm}\par}

{\vspace{2mm}\par\noindent {\it
$^{\dag}$~Department of Mathematics and Statistics, Memorial University of Newfoundland,\\
$\phantom{^\ddag}$St.\ John's (NL) A1C 5S7, Canada
}\par}

{\vspace{2mm}\par\noindent {\it
$^{\ddag}$~Fakult\"at f\"ur Mathematik, Universit\"at Wien, Oskar-Morgenstern-Platz 1, A-1090 Wien, Austria\\
$\phantom{^\ddag}$~Institute of Mathematics of NAS of Ukraine, 3 Tereshchenkivska Str., 01601 Kyiv, Ukraine
}\par}

\vspace{2mm}\par\noindent
\textup{E-mail:} ecardosobihlo@mun.ca, abihlo@mun.ca, rop@imath.kiev.ua

\vspace{4mm}\par\noindent\hspace*{5mm}\parbox{150mm}{\small
We find the complete equivalence group of a class of (1+1)-dimensional second-order evolution equations, which is infinite-dimensional. 
The equivariant moving frame methodology is invoked to construct, in the regular case of the normalization procedure, 
a moving frame for a group related to the equivalence group
in the context of equivalence transformations among equations of the class under consideration.
Using the moving frame constructed, we describe the algebra of differential invariants of the former group by obtaining 
a minimum generating set of differential invariants and a complete set of independent operators of invariant differentiation.
}\par\vspace{2mm}

\noprint{
MSC: 35K57 (primary), 35A30, 58A15 (secondary)
58-XX Global analysis, analysis on manifolds [See also 32Cxx, 32Fxx, 32Wxx, 46-XX, 47Hxx, 53Cxx, For geometric integration theory, see 49Q15]
 58Axx General theory of differentiable manifolds [See also 32Cxx]
  58A15 Exterior differential systems (Cartan theory)
35-XX Partial differential equations
 35Axx General topics
  35A30 Geometric theory, characteristics, transformations [See also 58J70, 58J72]
 35Kxx Parabolic equations and systems [See also 35Bxx, 35Dxx, 35R30, 35R35, 58J35]
  35K57 Reaction-diffusion equations
}

\section{Introduction}

Invariants and differential invariants of transformation groups, in particular, point symmetry groups admitted by systems of differential equations have a wide range of applications and are therefore an intensively investigated subject. Differential invariants play a central role in the invariant parameterization problem~\cite{bihl19a,bihl11Fy,popo10Cy} and in the problem of invariant discretization~\cite{bihl12By,bihl17b,doro11Ay}. They are also used to construct invariant differential equations and invariant variational problems~\cite{olve07Ay,olve11a}, as well as in computer vision, integrable systems, classical invariant theory and the calculus of variations~\cite{cheh08Ay,olve07Ay,olve15a}.

Rather recently, finding differential invariants in problems related to group classification became a research topic of interest. The idea is to compute the differential invariants not for the point symmetry group of a single system of differential equations but for the equivalence group admitted by a class of such systems. The primary motivation for such a survey is to study the equivalence of systems of differential equations. Exploring equivalence, it is possible to explicitly determine point transformations among systems from a class~\cite{ovsi82Ay}. Such a mapping between two systems of differential equations is especially helpful if for one of the systems involved wide sets of exact solutions are known. These solutions then can be mapped to solutions of the equivalent system. Another case of particular interest is the mapping between nonlinear and linear elements of a class of systems of differential equations~\cite{kume82Ay}. For the solution of the equivalence problem, finding differential invariants for the equivalence group is a main ingredient.
There are a number of papers where some low-order differential invariants of the equivalence groups of various physically relevant classes of systems of differential equations were computed using the Lie infinitesimal method; see, e.g., \cite{gand07Ay,ibra04By,ibra08Ay,ibra07Ay,ibra04Ay,john01Ay,torr05Ay,torr04Ay,trac04Ay,tsao09Ay} and references therein.

In the present paper we will be concerned with differential invariants for a group%
\footnote{In fact, this object and the ``equivalence group'' of the class~\eqref{eq:ClassOfDiffusionEquations} 
are Lie pseudogroups of locally defined point transformations. 
We use the term ``group'' for brevity since this does not lead to any confusion.} 
related to the equivalence group of the class of diffusion equations
\begin{equation}\label{eq:ClassOfDiffusionEquations}
 u_t=u_{xx}+f(u,u_x).
\end{equation}
in the context of equivalence transformations among equations of this class.
This subject was originally considered in~\cite{torr05Ay}, using the infinitesimal method and restricting the order of differential invariants up to two. 
We revisit the construction of differential invariants for the class~\eqref{eq:ClassOfDiffusionEquations} 
from the very beginning, analyzing differential invariants of which group should be found.
Then, we apply the method of equivariant moving frames in the formulation originally proposed and formulated by Fels and Olver~\cite{fels98Ay,fels99Ay}, 
which was later generalized to infinite-dimensional Lie (pseudo)groups in~\cite{cheh08Ay,olve08Ay,olve09Cy}, 
and this is the setting that is needed to study differential invariants for the class~\eqref{eq:ClassOfDiffusionEquations}. 
The advantage of moving frames is that they allow for a canonical process of \textit{invariantization}, which associates to each object, such as functions, differential functions, differential forms and total differentiation operators, its invariant counterpart. For the problem of finding differential invariants of a Lie transformation (pseudo)group, this property is especially convenient. The invariantization of the jet-space coordinate functions yields the so-called
\emph{normalized differential invariants}. The invariantized coordinate functions whose transformed counterparts were involved in the construction of the corresponding moving frame via the \emph{normalization procedure} are equal to the respective constants chosen in the course of normalization. This is why these objects are called \emph{phantom normalized differential invariants}. 
The \emph{non-phantom normalized differential invariants} constitute a complete set of functionally independent differential invariants. As a further asset, the method of moving frames also permits to study the algebra of differential invariants by deriving relations, called \textit{syzygies}, between invariant derivatives of non-phantom normalized differential invariants. Finding syzygies can aid in the establishment of a \emph{minimum generating set} of differential invariants. See e.g.~\cite{cheh08Ay,doss12a,olve07Ay,olve08Ay,olve09Cy} for more details and an extensive discussion on the computation of differential invariants for both finite-dimensional Lie symmetry groups and for infinite-dimensional Lie (pseudo)groups using moving frames.

The further organization of this paper is as follows. 
In Section~\ref{sec:EquivalenceGroupDiffusionEquations} we compute the equivalence group and the equivalence algebra of the class~\eqref{eq:ClassOfDiffusionEquations}. 
Section~\ref{sec:PreliminaryAnalysis} is devoted to the selection of a group to be considered and 
a preliminary analysis of equivariant moving frames associated with this group.
The structure of the algebra of differential invariants is determined in the main Section~\ref{sec:DiffInvsRegularCase}. 
This includes a description of a minimum generating set of differential invariants and a complete set of independent operators of invariant differentiation, 
which serve to exhaustively describe the set of differential invariants. 
Moreover, for each $k\in\mathbb N_0$ we explicitly present a functional basis of differential invariants 
of order not greater than~$k$.

\section{The equivalence group}\label{sec:EquivalenceGroupDiffusionEquations}

The auxiliary system for the class~\eqref{eq:ClassOfDiffusionEquations}, 
which is satisfied by the arbitrary element~$f$, is $f_t=f_x=f_{u_t}=f_{u_{tt}}=f_{u_{tx}}=f_{u_{xx}}=0$.
By definition~\cite{opan17a,ovsi82Ay,popo06Ay,popo10Ay}, 
the (usual) equivalence group~$G^\sim$ of the class~\eqref{eq:ClassOfDiffusionEquations} 
consists of the point transformations in the space with coordinates $(t,x,u,u_t,u_x,u_{tt},u_{tx},u_{xx},f)$
that have the following properties:
\begin{itemize}\itemsep=0ex
\item 
they are projectable to the space with the coordinates~$(t,x,u)$, 
\item 
their components for derivatives of~$u$ are found by prolongation using the chain rule, and  
\item 
they map every equation from the class~\eqref{eq:ClassOfDiffusionEquations} to an equation from the same class. 
\end{itemize}
To begin finding the group~$G^\sim$, we fix an arbitrary equation of the class~\eqref{eq:ClassOfDiffusionEquations}, $u_t=u_{xx}+f(u,u_x)$,
and aim to find point transformations in the space with coordinates $(t,x,u)$, 
\begin{equation}\label{eq:GeneralFormOfEquivalenceTransformation}
 \tilde t=T(t,x,u),\quad \tilde x=X(t,x,u),\quad \tilde u=U(t,x,u),
\end{equation}
that transform the fixed equation to an equation of the same class, 
\begin{equation}\label{eq:TransformedClassOfDiffusionEquations}
 \tilde u_{\tilde t}=\tilde u_{\tilde x\tilde x}+\tilde f(\tilde u,\tilde u_{\tilde x}).
\end{equation}
A preliminary simplification is obtained from noting that the class~\eqref{eq:ClassOfDiffusionEquations} 
is a subclass of the class of second-order (1+1)-dimensional semi-linear evolution equations. 
Any point transformation between two equations from the latter class satisfies the constraints $T_x=T_u=X_u=0$, 
i.e., $\tilde t=T(t)$, $\tilde x=X(t,x)$, and $T_tX_xU_u\ne0$. See~\cite{ivan10Ay,king98Ay,maga93Ay} for further details.
After taking into account the above constraints, the required transformed derivatives read
\[
 \tilde u_{\tilde t}=\frac1{T_t}\left(\mathrm D_tU-\frac{X_t}{X_x}\mathrm D_xU\right),\quad \tilde u_{\tilde x}=\frac1{X_x}\mathrm D_xU,\quad \tilde u_{\tilde x\tilde x}= \left(\frac1{X_x}\mathrm D_x\right)^2U,
\]
where $\mathrm D_t$ and $\mathrm D_x$ are the usual total derivative operators with respect to~$t$ and~$x$, respectively.
Substituting these expressions and $u_t=u_{xx}+f$
into Eq.~\eqref{eq:TransformedClassOfDiffusionEquations},  
we split the resulting equation with respect to $u_{xx}$ yielding $T_t=X_x^{\,2}$. 
The remaining equation is
\begin{gather}\label{eq:ReducedDetEq}
 f=\frac1{U_u}\left(T_t\tilde f-U_t+\frac{X_t}{X_x}(U_x+U_uu_x)+U_{xx}+2U_{xu}u_x+U_{uu}u_x^2\right).
\end{gather}
The differential consequences of Eq.~\eqref{eq:ReducedDetEq} that are obtained by separate differentiations with respect to~$t$ and~$x$ 
can be split with respect to derivatives of~$\tilde f$ since they are regarded as independent for equivalence transformations. 
This yields the equations $T_{tt}=X_{xt}=X_{tt}=U_t=U_x=0$. 
The equation~\eqref{eq:ReducedDetEq} itself gives the $f$-component of equivalence transformations.  

The arbitrary element~$f$ in fact depends only on~$u$ and~$u_x$. 
The space with coordinates $(t,x,u,u_x,f)$ is preserved by all elements of~$G^\sim$. 
This is why we can assume this space as the underlying space for~$G^\sim$ 
and present merely the transformation components for its coordinates.

As a result, we have proved the following theorem.

\begin{theorem}
The equivalence group~$G^\sim$ of the class~\eqref{eq:ClassOfDiffusionEquations} is constituted by the transformations
\begin{align}\label{eq:EquivalenceTransformationsDiffusionEquations}
\begin{split}
 &\tilde t=C_1^2t+C_0,\quad \tilde x=C_1x+C_1C_2t+C_3,\quad \tilde u=\varphi(u),\quad \tilde u_{\tilde x}=\frac{\varphi'}{C_1}u_x,\\
 &\tilde f=\frac1{C_1^2}\left(\varphi'f-C_2\varphi'u_x-\varphi''u_x^2\right),
\end{split}
\end{align}
where $C_0, C_1, C_2, C_3\in\mathbb{R}$, $\varphi$ is an arbitrary smooth function of $u$ and $C_1\varphi'\ne0$.
\end{theorem}
The infinitesimal generators of one-parameter subgroups of~$G^\sim$, 
which constitute the equivalence algebra~$\mathfrak g^\sim$ of the class~\eqref{eq:ClassOfDiffusionEquations}, 
can be derived from~\eqref{eq:EquivalenceTransformationsDiffusionEquations} by differentiation, 
cf.\ the proof of Corollary~11 in~\cite{kuru18a} or the proof of Corollary~6 in~\cite{bihl17c}. 
These generators coincide with those determined in~\cite{torr05Ay}. 
As we will later need them for the description of the algebra of differential invariants of a group related to~$G^\sim$ 
in the context of the $G^\sim$-equivalence among equations of the class~\eqref{eq:ClassOfDiffusionEquations}, 
we present them here. 
The general element of~$\mathfrak g^\sim$ is
\[
Q=\tau\p_t+\xi\p_x+\phi\p_u+\eta\p_{u_x}+\theta\p_f,
\]
where the components are of the form 
\begin{align*}
 &\tau=2c_1t+c_0,\quad \xi=c_1x+c_2t+c_3,\quad \phi=\phi(u),\\
 &\eta=(\phi'-c_1)u_x,\quad \theta=(\phi'-2c_1)f-c_2u_x-\phi''u_x^2,
\end{align*}
in which $c_0$, $c_1$, $c_2$ and $c_3$ are arbitrary real constants, 
and $\phi$ is an arbitrary smooth function of~$u$. 
In other words, the equivalence algebra~$\mathfrak g^\sim$ of the class~\eqref{eq:ClassOfDiffusionEquations} 
is spanned by the vector fields 
\begin{gather*}
\p_t,\quad
2t\p_t+x\p_x-u_x\p_{u_x}-2f\p_f,\quad
t\p_x-u_x\p_f,\quad
\phi\p_u+\phi'u_x\p_{u_x}+(\phi'f-\phi''u_x^2)\p_f,\quad
\end{gather*}
where $\phi$ runs through the set of smooth functions of $u$.

\section{Preliminary analysis of moving frames}\label{sec:PreliminaryAnalysis}

Let us first clarify the space of independent and dependent variables to be used and the group to be considered. 
While formally the arbitrary element~$f$ is a smooth function on the second-order jet space with coordinates $(t,x,u,u_t,u_x,u_{tt},u_{tx},u_{xx})$, 
practically it explicitly depends only on~$u$ and~$u_x$. 
This is why subsequently we will only consider the projection of the equivalence transformations to the space with coordinates $(u,u_x,f)$. 
As a shorthand, we denote $v:=u_x$ and $\tilde v:=\tilde u_{\tilde x}=V(u,v):=C_1^{-1}\varphi'(u)v$.
In other words, we will in fact study differential invariants of the projection~$G_1$ of~$G^\sim$ to the space with coordinates $(u,v,f)$, 
where $u$ and~$v$ are the independent variables and $f$ is the dependent variable. 
The infinitesimal counterpart of~$G_1$ is the projection~$\mathfrak g_1$ of~$\mathfrak g^\sim$ to the space with coordinates $(u,v,f)$. 

In order to describe the algebra of differential invariants of the group $G_1$, we now construct a moving frame for this group. 
Since it is infinite-dimensional, we have to use the machinery developed for Lie (pseudo)groups, 
see~\cite{cheh08Ay,olve08Ay} for an extensive description of this subject.

The first step in the construction of the moving frame is the computation of the lifted horizontal coframe, the dual of which yields the implicit total differentiation operators $\mathrm D_{\tilde u}$ and $\mathrm D_{\tilde v}$. 
For the equivalence transformations~\eqref{eq:EquivalenceTransformationsDiffusionEquations}, the lifted horizontal coframe is
\begin{align*}
 &\mathrm d_{\rm h}\tilde u=(\mathrm D_uU)\,\mathrm d u+(\mathrm D_vU)\,\mathrm d v=\varphi'\,\mathrm d u,\\
 &\mathrm d_{\rm h}\tilde v=(\mathrm D_uV)\,\mathrm d u+(\mathrm D_vV)\,\mathrm d v=\frac{\varphi''}{C_1}v\,\mathrm d u+\frac{\varphi'}{C_1}\,\mathrm d v.
\end{align*}
Computing the dual, we derive that 
\begin{equation}\label{eq:OperatorsOfTotalDifferentiationDiffusionEquations}
 \mathrm D_{\tilde u}=\frac1{\varphi'}\mathrm D_u-\frac{\varphi''}{(\varphi')^2}v\mathrm D_v,\quad \mathrm D_{\tilde v}=\frac{C_1}{\varphi'}\mathrm D_v
\end{equation}
are the required implicit differentiation operators. Acting with them on the transformation component for $f$, we find that
\[
\tilde f_{ij}=\frac{\p^{i+j}\tilde f}{\p\tilde u^i\p\tilde v^j}=\mathrm D_{\tilde u}^{\,\,i}\mathrm D_{\tilde v}^{\,\,j}F,
\]
where $i,j\in\mathbb N_0:=\mathbb N\cup\{0\}$ and $\tilde f_{00}=\tilde f=F:=C_1^{-2}(\varphi'f-C_2\varphi'v-\varphi''v^2)$ is the $f$-component of equivalence transformations. 
In particular, the derivatives up to order 2 are exhausted by
\begin{align*}
\tilde f_{10}={}&\frac1{C_1^2\varphi'}\left(\varphi'f_u+\varphi''(f-vf_v)-\varphi'''v^2+2\frac{(\varphi'')^2}{\varphi'}v^2\right),\\
\tilde f_{01}={}&\frac1{C_1\varphi'}\left(\varphi'f_v-C_2\varphi'-2\varphi''v\right),\\
\tilde f_{20}={}&\frac1{C_1^2\varphi'}\left(f_{uu}\!-\!\frac{\varphi''}{\varphi'}(f_u\!-\!2vf_{uv})\!+\!\left(\frac{\varphi''}{\varphi'}\right)^2\!\!v^2\!f_{vv}
\!+\!\bigg(\frac{\varphi''}{\varphi'}\bigg)'\!(f\!-\!vf_v)
\!-\!(\varphi')^2\!\left(\frac1{\varphi'}\left(\frac1{\varphi'}\right)''\right)'\!\!v^2\!\right)\!,\! \\
\tilde f_{11}={}&\frac1{C_1\varphi'^2}\left(\varphi'f_{uv}-\varphi''vf_{vv}-2\varphi'''v+4\frac{\varphi''^2}{\varphi'}v\right), \\
\tilde f_{02}={}&\frac1{\varphi'^2}(\varphi'f_{vv}-2\varphi'').
\end{align*}

There are a relative invariant and a relative conditional invariant which play a significant role in the following consideration. 
By taking the difference $\tilde f_{00}-\tilde v\tilde f_{01}$ we exclude the inessential constant $C_2$, which only arises in $\tilde f_{00}$ and $\tilde f_{01}$,
\[
 \tilde f_{00}-\tilde v\tilde f_{01}=\frac{1}{C_1^2}\left(\varphi'(f-vf_v)+\varphi''v^2\right).
\]
Combining further 
$2(\tilde f_{00}-\tilde v\tilde f_{01})+\tilde v^2\tilde f_{02}$ to exclude $\varphi''$, we obtain
\[
\tilde W=\frac{1}{C_1^2}W, \quad\mbox{where}\quad 
W=2f-2vf_v+v^2f_{vv}, \quad 
\tilde W=2\tilde f-2\tilde v\tilde f_{\tilde v}+\tilde v^2\tilde f_{\tilde v\tilde v},
\]
i.e., $W$ is a relative invariant of~$G_1$. 
In other words, the condition $W=0$ is preserved by any equivalence transformation in the class~\eqref{eq:ClassOfDiffusionEquations}.
Analogously, the combination $2\tilde f_{10}-v\tilde f_{11}$ gives 
\begin{gather}\label{eq:TransOfS}
 \tilde S=\frac1{C_1^2}S+\frac1{C_1^2}\frac{\varphi''}{\varphi'}W, \quad\mbox{where}\quad 
 S=2f_u-vf_{uv},\quad 
\tilde S=2\tilde f_{\tilde u}-\tilde v\tilde f_{\tilde u\tilde v}.
\end{gather}
This means that $S$ is a relative invariant of~$G_1$ if the condition $W=0$ is satisfied.
Values of the differential functions~$W$ and~$S$ determine which normalization conditions should be chosen. 

We next find appropriate normalization conditions, which form the basis for the construction of an equivariant moving frame. As~$\varphi$ arises only in~$U$, we can set $U$ to any value including zero. The value of $V$ can be set to any constant excluding zero, and all these possibilities are equivalent. We find it convenient to put $V=1$ and express $\varphi'=C_1/v$. 
The constraint $W=0$ singles out the \emph{singular} case for the moving frame construction, which has to be investigated separately. 
Within this singular case, there is the \emph{ultra-singular} subcase associated with the constraint $S=0$. 
Indeed, under the constraint $W=0$ the equation~\eqref{eq:TransOfS} can be solved for $C_1$ if and only if $S\ne0$.

\section{Differential invariants for the regular case}\label{sec:DiffInvsRegularCase}

In this paper, we only consider the \emph{regular} case for moving frames of~$G_1$, where $W\ne0$. 
In this case, the following normalization conditions can be used to determine a complete moving frame
\begin{equation}\label{eq:NormalizationMovingFrameDiffusionEquations}
\begin{split}
&\tilde u=0,\quad \tilde v=1,\quad \tilde f=1, \quad \tilde f_{01}=0,\quad \tilde f_{02}=0,\\ 
&\tilde f_{i0}=-\frac{v^2\varphi^{(i+2)}}{C_1^{\,\,2}(\varphi')^i}
+\frac1{C_1^{\,\,2}}\sum_{i'=0}^i\binom i{i'}\frac1{(\varphi')^{i'}}\left(\frac{\varphi''}{(\varphi')^2}\right)^{i-i'}f_{i',i-i'}
+\dots=0,\ i\in\mathbb N.
\end{split}
\end{equation}
In the expression for~$\tilde f_{i0}$, we presented only the summands with the highest-order derivatives of~$\varphi$ and~$f$, 
which are $\varphi^{(i+2)}$ and $f_{i',i-i'}$, $i'=0,\dots,i$, respectively. 
We solve the first five equations with respect to~$C_1$, $C_2$, $\varphi$, $\varphi'$ and $\varphi''$ 
and substitute the obtained expressions into the other equations. 
For each fixed~$i\in\mathbb N$, we solve the modified equation $\tilde f_{i0}=0$ 
in view of the similar equations with lower values of~$i$ and thus find an expression for~$\varphi^{(i+2)}$, 
the explicit form of which is essential for further consideration only for $i=3$. 
This yields the following complete moving frame:
\begin{align}\label{eq:MovingFrameDiffusionEquations}
\begin{split}
 &C_1=\frac W{2v},\quad C_2=f_v-vf_{vv},\quad \varphi=0,\quad \varphi'=\frac W{2v^2},\quad \varphi''=\frac W{4v^2}f_{vv},\\[.5ex]
 &\varphi'''=\frac W{4v^4}\Big(2f_u+(f-vf_v+v^2f_{vv})f_{vv}\Big), \\
 &\varphi^{(i+2)}=\frac W{2v^1}\sum_{i'=0}^i\binom i{i'}\left(\frac{v^2}W\right)^{i-i'}f_{i',i-i'}+\cdots,\quad i=2,3,\dots\,.
\end{split}
\end{align}
In the expression for~$\varphi^{(i+2)}$, we presented only the summands with the highest-order derivatives $f_{i',i-i'}$, $i'=0,\dots,i$. 
The invariantization $I^{ij}=\iota(f_{ij})$ of the derivatives~$f_{ij}$ of~$\tilde f$ 
that are not involved in the normalization conditions~\eqref{eq:NormalizationMovingFrameDiffusionEquations} 
gives rise to a complete set of functionally independent differential invariants of $G_1$. 
The lowest-order non-phantom normalized differential invariant is $I^{11}$, and it reads
\[
I^{11}=-2v^2\frac{4f_u-2vf_{uv}+(2f-2vf_v+v^2f_{vv})f_{vv}}{(2f-2vf_v+v^2f_{vv})^2}.
\]
This differential invariant is of second order.
For each tuple $(i,j)$ with $i+j\geqslant3$ and $j\ne0$, 
the maximal orders of derivatives of~$f$ and~$\varphi$ appearing in the expression for~$\tilde f_{ij}$ are $i+j$ and $i+2$, respectively. 
This is why the maximal order of derivatives of~$f$ in the expression for~$\tilde f_{ij}$ 
cannot be lowered in the course of the invariantization, 
i.e., the order of the normalized differential invariant~$I^{ij}$ is $i+j$.
Therefore, there are precisely $\frac12k(k+1)-2$ 
functionally independent differential $G_1$-invariants of order not greater than $k\geqslant2$. 
They are given by the functions $I^{11}$ and~$I^{ij}$ with $3\leqslant i+j\leqslant k$ and $j\ne0$. 

Apart from finding the complete set of functionally independent differential invariants of $G_1$ for each fixed order by successively invariantizing all the derivatives $f_{ij}$, the moving frame~\eqref{eq:MovingFrameDiffusionEquations} can be used to determine the operators of invariant differentiation. They are found upon invariantizing the operators of total differentiation~\eqref{eq:OperatorsOfTotalDifferentiationDiffusionEquations} and read
\begin{equation}\label{eq:OperatorsOfInvariantDifferentiationDiffusionEquations}
 \mathrm D_u^{\mathrm i}=\frac{2v^2}{2f-2vf_v+v^2f_{vv}}\left(\mathrm D_u-\frac12vf_{vv}\mathrm D_v\right),\quad \mathrm D_v^{\mathrm i}=v\mathrm D_v.
\end{equation}

We now aim to investigate the structure of the algebra of differential invariants of $G_1$. The starting point for this investigation is the universal recurrence relation, which relates the differentiated invariantized differential functions or differential forms with the invariantization of the respective differentiated objects. This universal recurrence relation reads~\cite{olve08Ay}
\begin{equation}\label{eq:UniversalRecurrenceRelation}
\mathrm d \iota (\Omega)=\iota\big(\mathrm d \Omega+Q^{(\infty)}(\Omega)\big).
\end{equation}
The first step in our study is the evaluation of~\eqref{eq:UniversalRecurrenceRelation} for the independent variables~$u$ and~$v$ and the derivatives $f_{ij}$, $i,j\in\mathbb N_0$,
\begin{gather*}
\mathrm d_{\rm h}\iota(u)=\omega^1+\iota(\phi), \quad
\mathrm d_{\rm h}\iota(v)=\omega^2+\iota(\eta), \\
\mathrm d_{\rm h}I^{ij}=\mathrm d_{\rm h}\iota(f_{ij})=\iota(f_{i+1,j}\mathrm d u+f_{i,j+1}\mathrm d v+\theta^{ij})=I^{i+1,j}\omega^1+I^{i,j+1}\omega^2+\iota(\theta^{ij}),
\end{gather*}
where $\omega^1=\iota(\mathrm d u)$, $\omega^2=\iota(\mathrm d v)$, and 
\[
\begin{split}
\theta^{ij}={}&\mathrm D_u^{\,\,i}\mathrm D_v^{\,\,j}(\theta-\phi f_{10}-\eta f_{01})+\phi f_{i+1,j}+\eta f_{i,j+1}\\
={}&-(j-1)\sum_{i'=0}^i\binom i{i'}\phi^{(i'+1)}f_{i-i',j}
-\sum_{i'=1}^i\binom i{i'}\Big(\phi^{(i')}f_{i-i'+1,j}+v\phi^{(i'+1)}f_{i-i',j+1}\Big)\\
&{}+(j-2)c_1f_{ij}-c_2\delta_{0i}(\delta_{0j}v+\delta_{1j})-\phi^{(i+2)}(\delta_{0j}v^2+2\delta_{1j}v+2\delta_{2j})
\end{split}
\]
is the $f_{ij}$-component of the infinite prolongation of the vector field~$\phi\p_u+\eta\p_v+\theta\p_f$. 
Here $\delta_{ij}$ is the Kronecker delta. 
The respective recurrence relations then split into two kinds, the first being the so-called phantom recurrence relations. For a well-defined moving frame cross-section, they can be uniquely solved for the invariantized Maurer--Cartan forms, which arise due to the presence of the correction term $\iota\big(Q^{(\infty)}(\Omega)\big)$ in~\eqref{eq:UniversalRecurrenceRelation}. Then, plugging these invariantized Maurer--Cartan forms into the second kind of recurrence relations, the non-phantom ones, gives a complete description of the relation between the normalized and differentiated differential invariants, see~\cite{cheh08Ay,olve08Ay} for more details. For the chosen cross-section~\eqref{eq:NormalizationMovingFrameDiffusionEquations}, the phantom recurrence relations read
\begin{align*}
 0&=\mathrm d_{\rm h}\iota(u)= \omega^1+\iota(\phi)=\omega^1+\hat\phi,\\ 
 0&=\mathrm d_{\rm h}\iota(v)= \omega^2+\iota(\eta)=\omega^2+\hat\phi'-\hat c_1,\\
 0&=\mathrm d_{\rm h} I^{00} = \iota(\theta)=\hat\phi'-2\hat c_1-\hat c_2-\hat\phi'',\\
 0&=\mathrm d_{\rm h} I^{01} = I^{11}\omega^1+\iota(\theta^{01})=I^{11}\omega^1-\hat c_2-2\hat\phi'',\\
 0&=\mathrm d_{\rm h} I^{02} = I^{12}\omega^1+I^{03}\omega^2+\iota(\theta^{02})=I^{12}\omega^1+I^{03}\omega^2-2\hat\phi'',\\
 0&=\mathrm d_{\rm h} I^{i0} = I^{i1}\omega^2+\iota(\theta^{i0}) \\[-1.3ex]
  &=I^{i1}\omega^2+\hat\phi^{(i+1)}-\hat\phi^{(i+2)}-\sum_{i'=1}^{i-1}\binom i{i'}I^{i-i',1}\hat\phi^{(i'+1)},\quad i\in\mathbb N,
\end{align*}
where the forms $\hat c_1$, $\hat c_2$ and \smash{$\hat\phi^{(i)}$}, $i\in\mathbb N_0$, 
are the invariantizations of the parameters~$c_1$, $c_2$ and~$\phi^{(i)}$ of the infinitely prolonged general element 
of the projected algebra~$\mathfrak g_1$, respectively, 
$\hat c_1=\iota(c_1)$, $\hat c_2=\iota(c_2)$ and $\hat\phi^{(i)}=\iota(\phi^{(i)})$.
More rigorously, here the parameters~$c_1$, $c_2$ and~$\phi^{(i)}$, $i\in\mathbb N_0$, are interpreted 
as the coordinate functions on the infinite prolongation of~$\mathfrak g_1$. 
Recall that under the prolongation we consider $u$ and~$v$ to be the independent variables and $f$ to be the dependent variable. 
In other words, these coefficients are first-order differential forms in the jet space $\mathrm J^\infty(u,v|\,f)$.
Hence their invariantizations are also forms, which are called \emph{invariantized Maurer--Cartan forms}.

The above system can be solved to yield the following invariantized Maurer--Cartan forms
\begin{align}\label{eq:InvariantizedMaurerCartanFormsDiffusionEquations}
\begin{split}
&\hat c_1=\left(\frac12I^{12}-I^{11}\right)\omega^1+\left(\frac12I^{03}-1\right)\omega^2,\quad
 \hat c_2=(I^{11}-I^{12})\omega^1-I^{03}\omega^2,\\
&\hat\phi=-\omega^1,\quad
\hat\phi'=\left(\frac12I^{12}-I^{11}\right)\omega^1+\left(\frac12I^{03}-2\right)\omega^2,\quad
\hat\phi''=\frac12I^{12}\omega^1+\frac12I^{03}\omega^2,\\   
& \hat\phi^{(i+2)}=\hat\phi^{(i+1)}-\sum_{i'=1}^{i-1}\binom i{i'}I^{i-i',1}\hat\phi^{(i'+1)}+I^{i1}\omega^2,\quad i\in\mathbb N.
\end{split}
\end{align}
The explicit expression for the invariantized form~$\hat\phi^{(i+2)}$, $i\in\mathbb N$, as a combination of~$\omega^1$ and~$\omega^2$ 
with coefficients being polynomials of normalized differential invariants is obtained by expanding the above expression
when successively going over the values of~$i$.  
In particular, 
\begin{gather*}
\hat\phi'''=\frac12I^{12}\omega^1+\left(I^{11}+\frac12I^{03}\right)\omega^2,\\
\hat\phi^{(4)}=\left(\frac12I^{12}-I^{11}I^{12}\right)\omega^1+\left(I^{11}+\frac12I^{03}+I^{21}-I^{11}I^{03}\right)\omega^2.
\end{gather*}
For $i\geqslant3$, the greatest value of $i'+j'$ for the normalized differential invariants~$I^{i'j'}$ 
that are involved in~$\hat\phi^{(i+2)}$ is $i+1$, 
and $I^{i1}\omega^2$ is the only summand with this value. 

The non-phantom recurrence relations are
\begin{align*}
\mathrm d_{\rm h} I^{11} &= I^{21}\omega^1+I^{12}\omega^2+\iota(\theta^{11})\\
 &{}=(I^{21}+2(I^{11})^2-I^{11}I^{12}-I^{12})\omega^1+(I^{12}-I^{11}I^{03}+I^{11}-I^{03})\omega^2,\\[1ex] 
\mathrm d_{\rm h} I^{ij} &= I^{i+1,j}\omega^1+I^{i,j+1}\omega^2+\iota(\theta^{ij}),\quad i+j\geqslant3,\ j\ne0,
\end{align*}
with\vspace*{-1ex}
\[
\begin{split}
\iota(\theta^{ij})={}&-(j-1)\sum_{i'=0}^i\binom i{i'}I^{i-i',j}\hat\phi^{(i'+1)}
-\sum_{i'=1}^i\binom i{i'}\Big(I^{i-i'+1,j}\hat\phi^{(i')}+I^{i-i',j+1}\hat\phi^{(i'+1)}\Big)\\
&{}+(j-2)I^{ij}\hat c_1-\delta_{0i}(\delta_{0j}+\delta_{1j})\hat c_2-(\delta_{0j}+2\delta_{1j}+2\delta_{2j})\hat\phi^{(i+2)}.
\end{split}
\]
The first non-phantom recurrence relation splits into 
\[
\mathrm D_u^{\mathrm i}I^{11}=I^{21}+2(I^{11})^2-I^{11}I^{12}-I^{12},\quad \mathrm D_v^{\mathrm i}I^{11}=I^{12}-I^{11}I^{03}+I^{11}-I^{03}.
\]
Therefore, the normalized differential invariants~$I^{12}$ and~$I^{21}$ are expressed in terms of invariant derivatives of~$I^{11}$ and~$I^{03}$, 
\begin{gather}\label{eq:ExpressionsForI12AndI21}
\begin{split}
&I^{12}=\mathrm D_v^{\mathrm i}I^{11}+I^{11}I^{03}-I^{11}+I^{03},\\ 
&I^{21}=\mathrm D_u^{\mathrm i}I^{11}-2(I^{11})^2+(I^{11}+1)(\mathrm D_v^{\mathrm i}I^{11}+I^{11}I^{03}-I^{11}+I^{03}).
\end{split}
\end{gather}
In view of the above discussion on the invariantize forms~$\hat\phi^{(i')}$, $i\in\mathbb N$, 
the expression for $\iota(\theta^{ij})$ with $i+j\geqslant3$ and $j\ne0$ implies that 
the greatest value of $i'+j'$ for~$I^{i'j'}$ involved in~$\iota(\theta^{ij})$ is $i+j$. 
Hence splitting the recurrence relation with $\mathrm d_{\rm h} I^{ij}$ 
leads to expressions for~$I^{i+1,j}$ and~$I^{i,j+1}$ in terms of invariant derivatives of~$I^{i'j'}$ with $i'+j'\leqslant i+j$. 
For example, from the non-phantom recurrence relation
\begin{align*}
\mathrm d_{\rm h} I^{03} &= I^{13}\omega^1+I^{04}\omega^2+\iota(\theta^{03})\\
 &{}=\left(I^{13}+I^{11}I^{03}-\frac12I^{12}I^{03}\right)\omega^1+ \left(I^{04}+I^{03}-\frac12(I^{03})^2\right)\omega^2
\end{align*}
we derive 
\begin{align*}
I^{13}=\mathrm D_u^{\mathrm i}I^{03}-I^{11}I^{03}+\frac12I^{12}I^{03},\quad 
I^{04}=\mathrm D_v^{\mathrm i}I^{03}+\frac12(I^{03})^2-I^{03}.
\end{align*}
This implies by induction, where the expressions~\eqref{eq:ExpressionsForI12AndI21} for~$I^{12}$ and~$I^{21}$ give the base case, 
that any non-phantom normalized differential invariant can be expressed in terms of invariant derivatives of~$I^{11}$ and~$I^{03}$. 

To find a minimum generating set of differential invariants for the projected group $G_1$, 
we should additionally check whether $I^{03}$ can be expressed in terms of invariant derivatives of~$I^{11}$.
We use~\eqref{eq:UniversalRecurrenceRelation} to compute the commutator between the operators of invariant differentiation. 
This is done upon evaluating~\eqref{eq:UniversalRecurrenceRelation} for the basis horizontal forms $\mathrm d u$ and $\mathrm d v$,
\begin{align*}
&\mathrm d_{\rm h} \iota(\mathrm d u) = \iota (\phi'\mathrm d u)=\iota(\phi')\wedge\iota(\mathrm d u)
=\left(2-\frac12I^{03}\right)\omega^1\wedge\omega^2=-Y^1_{12}\,\omega^1\wedge\omega^2,\\
&\mathrm d_{\rm h} \iota(\mathrm d v) = \iota\big(\phi''v\mathrm d u+(\phi'-c_1)\mathrm dv\big)
=\iota(\phi''v)\wedge\iota(\mathrm d u)=-\frac12I^{03}\omega^1\wedge\omega^2=-Y^2_{12}\,\omega^1\wedge\omega^2.
\end{align*}
The commutation relation then evaluates as
\[
 [\mathrm D_u^{\mathrm i},\mathrm D_v^{\mathrm i}]=Y^1_{12}\mathrm D_u^{\mathrm i}+Y^2_{12}\mathrm D_v^{\mathrm i}=\left(\frac12I^{03}-2\right)\mathrm D_u^{\mathrm i}+\frac12I^{03}\mathrm D_v^{\mathrm i}, 
\]
see~\cite{olve08Ay} for details of the technique applied. 
Evaluating $[\mathrm D_u^{\mathrm i},\mathrm D_v^{\mathrm i}]I^{11}$, we can derive the following expression for~$I^{03}$:
\[
I^{03}:=\frac{2v^3f_{vvv}}{2f-2vf_v+v^2f_{vv}}
=2\frac{2\mathrm D_u^{\mathrm i}I^{11}+[\mathrm D_u^{\mathrm i},\mathrm D_v^{\mathrm i}]I^{11}}{\mathrm D_u^{\mathrm i}I^{11}+\mathrm D_v^{\mathrm i}I^{11}}.
\]
As a result, we have proved the following theorem.

\begin{theorem}
The algebra of differential invariants of the group~$G_1$, 
which is the projection of the equivalence group $G^\sim$ 
of the class of diffusion equations~\eqref{eq:ClassOfDiffusionEquations} to the space with coordinates $(u,v,f)$, 
is generated by the single differential invariant 
\[
I^{11}=-2v^2\frac{4f_u-2vf_{uv}+(2f-2vf_v+v^2f_{vv})f_{vv}}{(2f-2vf_v+v^2f_{vv})^2}
\]
along with the two operators of invariant differentiation
\[
 \mathrm D_u^{\mathrm i}=\frac{2v^2}{2f-2vf_v+v^2f_{vv}}\left(\mathrm D_u-\frac12vf_{vv}\mathrm D_v\right),\quad \mathrm D_v^{\mathrm i}=v\mathrm D_v.
\]
All other differential invariants are functions of $I^{11}$ and invariant derivatives thereof.
\end{theorem}

\begin{corollary}
A functional basis of differential invariants of order not greater than~$k\in\mathbb N_0$ 
in terms of invariant derivatives of non-phantom normalized differential invariants is exhausted by 
\[
\big(\mathrm D_u^{\mathrm i}\big)^i\big(\mathrm D_v^{\mathrm i}\big)^jI^{11},\ i+j\leqslant k-2,\quad 
\big(\mathrm D_v^{\mathrm i}\big)^{j'}I^{03},\ j'\leqslant k-3.
\]
\end{corollary}

\section*{Acknowledgements}

This research was undertaken, in part, thanks to funding from the Canada Research Chairs program, the NSERC Discovery Grant program and the InnovateNL LeverageR{\&}D program.
AB is a recipient of an APART Fellowship of the Austrian Academy of Sciences.
The research of ROP and EMDSCB was supported by the Austrian Science Fund (FWF), projects P25064 and P29177. 

\footnotesize

\begin{thebibliography}{10}

\bibitem{bihl19a}
Bihlo A., Dos Santos Cardoso-Bihlo E. and Popovych R.O., 
Invariant parameterization of geostrophic eddies in the ocean, 
2019, arXiv:1908.06345, 21~pp.

\bibitem{bihl11Fy}
Bihlo A., Dos Santos Cardoso-Bihlo E.M. and Popovych R.O., 
Invariant parameterization and turbulence modeling on the beta-plane, 
\emph{Phys. D} \textbf{269} (2014), 48--62, arXiv:1112.1917.

\bibitem{bihl12By}
Bihlo A. and Popovych R.O., 
Invariant discretization schemes for the shallow-water equations, 
\emph{SIAM J. Sci. Comput.} \textbf{34} (2012), B810--B839, arXiv:1201.0498.

\bibitem{bihl17c}
Bihlo A. and Popovych R.O., 
Group classification of linear evolution equations,
\emph{J. Math. Anal. Appl.} \textbf{448} (2017), 982--1005, arXiv:1605.09251.

\bibitem{bihl17b}
Bihlo A. and Valiquette F., 
Symmetry-preserving numerical schemes, 
in \emph{Symmetries and Integrability of Difference Equations}, Springer, 2017, pp.~261--324.

\bibitem{cheh08Ay}
Cheh J., Olver P.J. and Pohjanpelto J., 
Algorithms for differential invariants of symmetry groups of differential equations, 
\emph{Found. Comput. Math.} \textbf{8} (2008), 501--532.

\bibitem{doro11Ay}
Dorodnitsyn V., 
\emph{Applications of Lie Groups to Difference Equations},
vol.~8 of \emph{Differential and integral equations and their applications}, Chapman \& Hall/CRC, Boca Raton, FL, 2011.

\bibitem{doss12a}
Dos Santos Cardoso-Bihlo E.M., 
Differential invariants for the Korteweg--de Vries equation, 
in \emph{Proceedings of 6th International Workshop ``Group Analysis of Differential Equations \& Integrable Systems'' (June 17--21, 2012, Protaras, Cyprus)}, University of Cyprus, Nicosia, 2013, pp. 71--79.

\bibitem{fels98Ay}
Fels M. and Olver P.J., 
Moving coframes: I. A practical algorithm, 
\emph{Acta Appl. Math.} \textbf{51} (1998), 161--213.

\bibitem{fels99Ay}
Fels M. and Olver P.J., 
Moving coframes. II. Regularization and theoretical foundations, 
\emph{Acta Appl. Math.} \textbf{55} (1999), 127--208.

\bibitem{gand07Ay}
Gandarias M.L., Torrisi M. and Tracina R., 
On some differential invariants for a family of diffusion equations, 
\emph{J.~Phys.~A} \textbf{40} (2007), 8803--8813.

\bibitem{ibra04By}
Ibragimov N.H., 
Invariants of hyperbolic equations: Solution of the Laplace problem, 
\emph{J. Appl. Mech. Tech. Phys.} \textbf{45} (2004), 158--166.

\bibitem{ibra08Ay}
Ibragimov N.H., Meleshko S.V. and Thailert E., 
Invariants of linear parabolic differential equations, \emph{Commun. Nonlinear Sci. Numer. Simul.} \textbf{13} (2008), 277--284.

\bibitem{ibra07Ay}
Ibragimov N.H. and Sophocleous C., 
Differential invariants of the one-dimensional quasi-linear second-order evolution equation, 
\emph{Commun. Nonlinear Sci. Numer. Simul.} \textbf{12} (2007), 1133--1145.

\bibitem{ibra04Ay}
Ibragimov N.H., Torrisi M. and Valenti A., 
Differential invariants of nonlinear equations $v_{tt}= f (x, v_x) v_{xx}+ g (x, v_x)$, 
\emph{Commun. Nonlinear Sci. Numer. Simul.} \textbf{9} (2004), 69--80.

\bibitem{ivan10Ay}
Ivanova N.M., Popovych R.O. and Sophocleous C., 
Group analysis of variable coefficient diffusion-convection equations. I. Enhanced group classification, 
\emph{Lobachevskii J. Math.} \textbf{31} (2010), 100--122.

\bibitem{john01Ay}
Johnpillai I.K. and Mahomed F.M., 
Singular invariant equation for the $(1+1)$ Fokker--Planck equation, 
\emph{J.~Phys.~A} \textbf{34} (2001), 11033--11051.

\bibitem{king98Ay}
Kingston J.G. and Sophocleous C., 
On form-preserving point transformations of partial differential equations, 
\emph{J. Phys. A} \textbf{31} (1998), 1597--1619.

\bibitem{kume82Ay}
Kumei S. and Bluman G.W., 
When nonlinear differential equations are equivalent to linear differential equations, 
\emph{SIAM J. Appl. Math.} \textbf{42} (1982), 1157--1173.

\bibitem{kuru18a}
Kurujyibwami C., Basarab-Horwath P. and Popovych R.O., 
Algebraic method for group classification of (1+1)-dimensional linear Schr\"{o}dinger equations, 
\emph{Acta Appl. Math.} \textbf{157} (2018), 171--203, arXiv:1607.04118.

\bibitem{maga93Ay}
Magadeev B.A., 
Group classification of nonlinear evolution equations,
\emph{Algebra i Analiz} \textbf{5} (1993), 141--156, (in Russian); English translation in {\it St.~Petersburg Math.~J.} {\bf 5} (1994), 345--359.

\bibitem{olve07Ay}
Olver P.J., 
Generating differential invariants, 
\emph{J. Math. Anal. Appl.} \textbf{333} (2007), 450--471.

\bibitem{olve11a}
Olver P.J., 
Differential invariant algebra, 
\emph{Comtemp. Math.} \textbf{549} (2011), 95--121.

\bibitem{olve15a}
Olver P.J., 
Modern developments in the theory and applications of moving frames, 
\emph{London Math. Soc. Impact150 Stories} \textbf{1} (2015), 14--50.

\bibitem{olve08Ay}
Olver P.J. and Pohjanpelto J., 
Moving frames for {L}ie pseudo-groups, \emph{Canadian J. Math.} \textbf{60} (2008), 1336--1386.

\bibitem{olve09Cy}
Olver P.J. and Pohjanpelto J., 
Differential invariant algebras of Lie pseudo-groups, 
\emph{Adv. Math.} \textbf{222} (2009), 1746--1792.

\bibitem{opan17a}
Opanasenko S., Bihlo A. and Popovych R.O., 
Group analysis of general Burgers--Korteweg--de Vries equations, 
\emph{J. Math. Phys.} \textbf{58} (2017), 081511, 37~pp., arXiv:1703.06932.

\bibitem{ovsi82Ay}
Ovsiannikov L.V., 
Group analysis of differential equations, Acad. Press, New York, 1982.

\bibitem{popo06Ay}
Popovych R.O., 
Classification of admissible transformations of differential equations, 
in \emph{Collection of Works of Institute of Mathematics}, vol.~3, Institute of Mathematics, Kyiv, 2006, pp. 239--254.

\bibitem{popo10Cy}
Popovych R.O. and Bihlo A., 
Symmetry preserving parameterization schemes,
\emph{J.~Math. Phys.} \textbf{53} (2012), 073102 (36 pages), arXiv:1010.3010.

\bibitem{popo10Ay}
Popovych R.O., Kunzinger M. and Eshraghi H., 
Admissible transformations and normalized classes of nonlinear Schr\"odinger equations, 
\emph{Acta Appl. Math.} \textbf{109} (2010), 315--359, arXiv:math-ph/0611061.

\bibitem{torr05Ay}
Torrisi M. and Tracina R., 
Second-order differential invariants of a family of diffusion equations, 
\emph{J.~Phys.~A} \textbf{38} (2005), 7519--7526.

\bibitem{torr04Ay}
Torrisi M., Tracina R. and Valenti A., 
On the linearization of semilinear wave equations, 
\emph{Nonlin. Dyn.} \textbf{36} (2004), 97--106.

\bibitem{trac04Ay}
Tracina R., 
Invariants of a family of nonlinear wave equations, 
\emph{Commun. Nonlinear Sci. Numer. Simul.} \textbf{9} (2004), 127--133.

\bibitem{tsao09Ay}
Tsaousi C., Sophocleous C. and Tracina R., 
Invariants of two-and three-dimensional hyperbolic equations, 
\emph{J.~Math. Anal. Appl.} \textbf{349} (2009), 516--525.

\end{thebibliography}

\end{document}